\documentclass[conference]{IEEEtran}%
\usepackage{amsmath}
\usepackage{amsthm}
\usepackage{pgfplots}
\usepackage{amsfonts}
\usepackage{amssymb}
\usepackage{graphicx}
\usepackage{algorithmic}
\usepackage{algorithm}
\usepackage{todonotes}
\usepackage{mathtools}
\usepackage{manfnt}
\usepackage{balance}%
\providecommand{\U}[1]{\protect\rule{.1in}{.1in}}

\theoremstyle{definition}

\newcommand{\CS}[1]{{#1}}
\newenvironment{figure2}{\begin{figure*}}{\end{figure*}}
\begin{document}

\title{Compressive Coded Random Access for Massive MTC Traffic in 5G Systems}
\author{Gerhard Wunder$^{1}$, \v Cedomir Stefanovi\' c$^{2}$, Petar Popovski$^{2}$, Lars Thiele$^{3}$\\
$^{1}$Heisenberg Communications and Information Theory Group, Freie
Universit\"{a}t Berlin\\
$^{2}$Department of Electronic Systems, Aalborg University, Denmark\\
$^{3}$Fraunhofer Heinrich Hertz Institute}

\maketitle

\begin{abstract}
Massive MTC support is an important future market segment, but not yet
efficiently supported in cellular systems. In this paper we follow-up on
recent concepts combining advanced MAC protocols with Compressed Sensing (CS)
based multiuser detection. Specifically, we introduce a concept for sparse
joint activity, channel and data detection in the context of the Coded ALOHA
(FDMA) protocol. We will argue that a simple sparse activity and data
detection is not sufficient (as many papers do) because control resources are
in the order of the data. In addition, we will improve on the performance of
such protocols in terms of the reduction of resources required for the user
activity, channel estimation and data detection. We will mathematically
analyze the system accordingly and provide expressions for the capture
probabilities of the underlying sparse multiuser detector. Finally, we will
provide 'structured' CS algorithms for the joint estimation scheme and
evaluate its performance.

\end{abstract}

\let\thefootnote\relax\footnote{This work was carried out within DFG grants WU
598/7-1 and WU 598/8-1 (DFG Priority Program on Compressed Sensing). Part of
this work has been performed in the framework of the Horizon 2020 project
FANTASTIC-5G (ICT-671660), which is partly funded by the European Union. The
authors would like to acknowledge the contributions of their colleagues in
FANTASTIC-5G. The work of \v Cedomir Stefanovi\' c was supported by the Danish
for Independent Research, grant no. DFF-4005-00281.}




\section{Introduction}

The Internet of Things (IoT) is a most promising 5G market segment and in the
focus of all key players in the ICT domain. Even pessimistic forecasts predict
several billions of connected devices. Major proliferation of the IoT will be
naturally in the 5G wireless domain. Currently, IoT market is mainly served by
short range capillary wireless technologies such as Bluetooth LE, ZigBee, and
WiFi and proprietary (clean slate) low power wide area technologies such as
SIGFOX, LoRA etc. There is only small share for cellular and there is clearly
a need to act fast in this direction.

\begin{figure}[t]
\centering\includegraphics[width=1\linewidth]{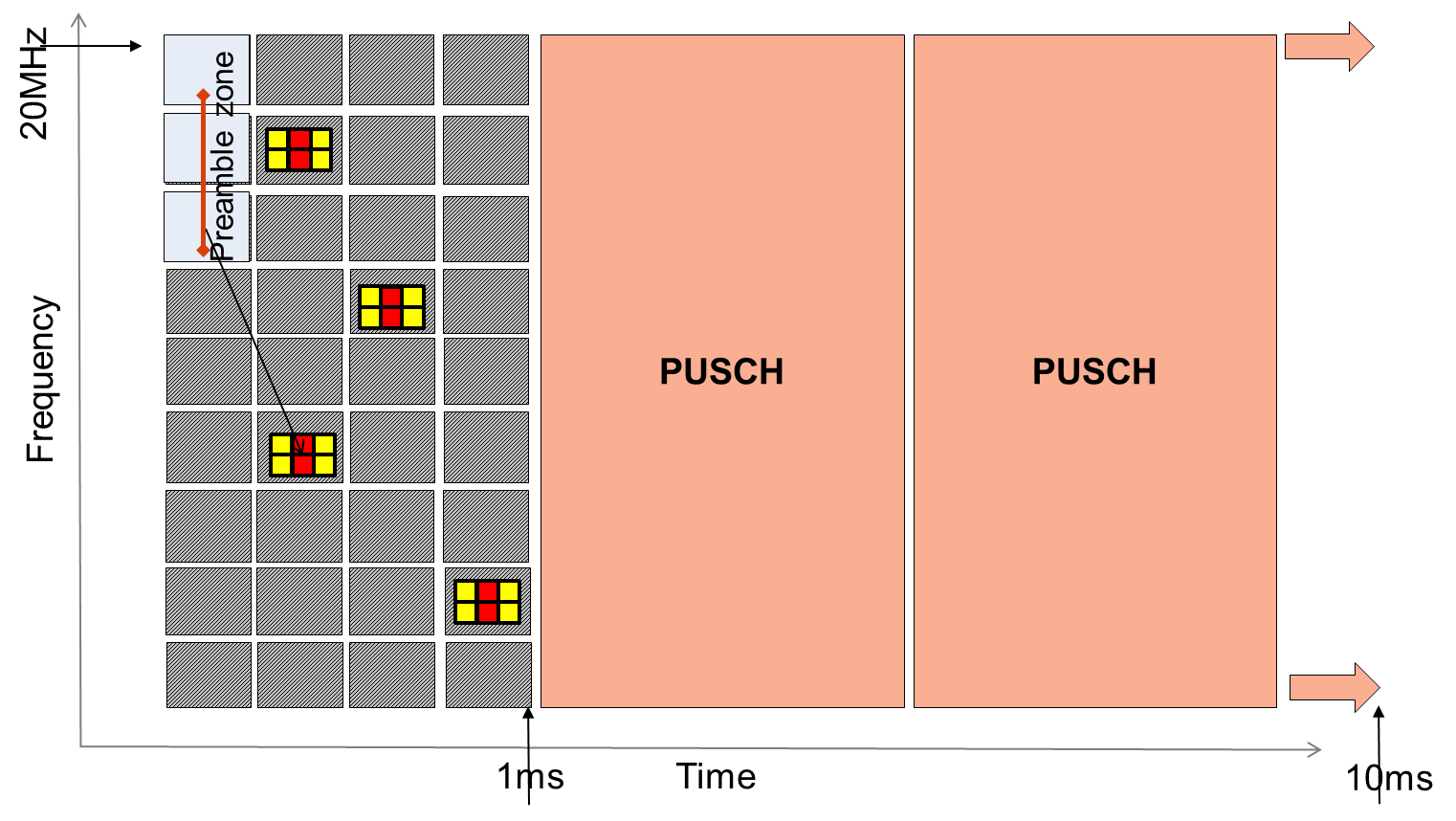}\caption{Random
access concepts: 5G standard approach with separated preamble section for
"activity" detection and respective pointer to data packet.}%
\label{fig:traditionell}%
\end{figure}

IoT requires support of scalable massive machine-type communication (MTC),
which is essentially a sporadic traffic pattern generated by devices operating
under tight (resource) constraints such as low cost, battery lifetime,
computation capability etc. Such messages have typically very unfavorable
control/data signaling ratio; recent proposals suggest 5G \textquotedblleft
one-shot\textquotedblright\ random access concepts where devices wake up and
send data right away with no coordination whatsoever
\cite{Wunder2014_COMMAG,Wunder2015_ACCESS}. The concept is depicted in Fig.
\ref{fig:traditionell}. While this concept is quite appealing it comes with
significant challenges:

\begin{itemize}
\item[(i)] Temporal asynchronous access among different resources; spectral
asynchronous access due to low-cost terminals; definition of shorter TTIs and
more granularity in allocating the physical resource blocks. \emph{This is the
waveform challenge} \cite{Wunder2014_COMMAG}.

\item[(ii)] Relationship between the data and the control data (metadata);
control signaling possibly in the order of data; per user resource control
signaling becomes inefficient. \emph{This is the metadata challenge}
\cite{Wunder2014_COMMAG}.

\item[(iii)] Throughput severely degraded due to collisions in random access
unless successive cancellation is applied. \emph{This is the throughput
challenge} \cite{PSLP2015}.
\end{itemize}

\begin{figure}[t]
\centering\includegraphics[width=1\linewidth]{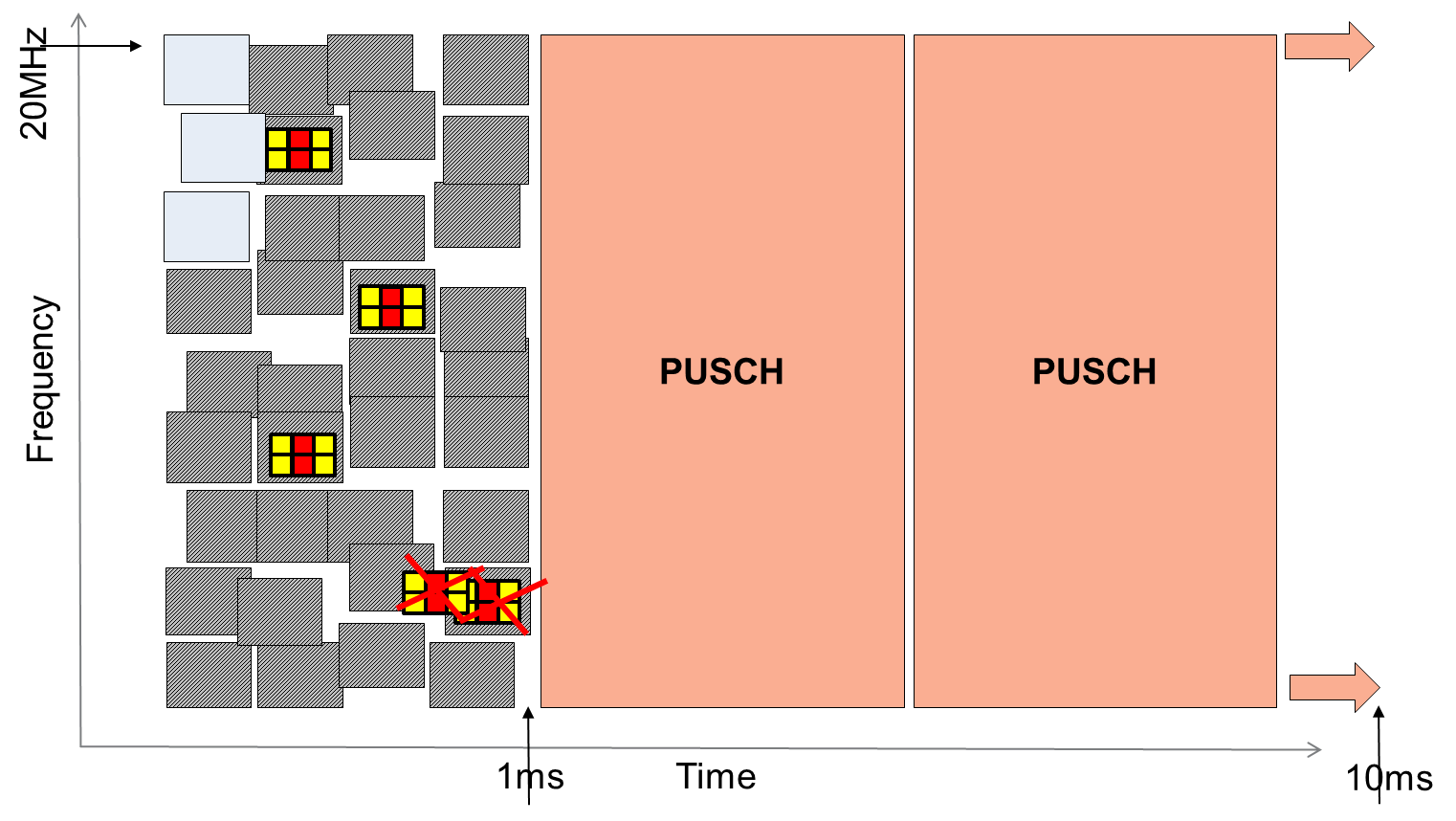}\caption{Random
access concepts: Traditonal approach.}%
\label{fig:challenges}%
\end{figure}

The challenges are depicted in Fig. \ref{fig:challenges}. In this paper, we
address the throughput challenge and follow-up on recent concepts combining
advanced MAC protocols with Compressed Sensing (CS) based multiuser detection.
Specifically, we introduce a concept for sparse joint activity, channel and
data detection in the context of the Coded ALOHA (FDMA) protocol which we call
\emph{Compressive Coded Random Access} (CCRA) extending the work in
\cite{Popovski14,Wunder2014_ICC,Wunder2015_GC}. We will argue that a simple
sparse activity and data detection is not sufficient (as many papers do)
because control resources are in the order of the data. In addition, we will
improve on the performance of such protocols in terms of the reduction of
resources required for the user activity, channel estimation and data
detection. We will mathematically analyze the system accordingly and provide
expressions for the capture probabilities of the underlying sparse multiuser
detector. Finally, we will provide 'structured' CS algorithms for the joint
estimation scheme and evaluate its performance.

\textbf{Notations}. $\lVert x\rVert_{\ell_{q}}=(\sum_{i}|x_{i}|^{q})^{1/q}$ is
the usual notion of $\ell_{q}$-norms and $\lVert x\rVert:=\lVert x\rVert
_{\ell_{2}}$. We denote with $\text{supp}(x):=\{i\,:\,x_{i}:=\langle
e_{i},x\rangle\neq0\}$ the support of $x$ in a given fixed (here canonical)
basis $\{e_{i}\}_{i=1}^{n}$. The size of its support is denoted as $\lVert
x\rVert_{\ell_{0}}:=|\text{supp}(x)|$. $W$ is the (unitary) Fourier matrix
with elements $(W)_{ij}=n^{-\frac{1}{2}}e^{-\sqrt{-1}2\pi ij/n}$ for
$k,l=0\dots n-1$, hence, $W^{-1}=W^{\ast}$ where $W^{\ast}$ is the adjoint of
$W$. We use $\hat{x}=Wx$ to denote Fourier transforms and $\odot$ means
point-wise product. $I_{n}$ is the identity matrix in $\mathbb{C}^{n}$,
diag$(x)$ is some arbitrary diagonal matrix with $x\in\mathbb{C}^{n}$ on its diagonal.

\section{CCRA model}

\label{sec:CCRA}

For simplicity assume one time slot only and $n$ OFDM subcarriers. This is
easily generalized to the case where there are multiple time slots, notably,
within the coherence time so that channels are constant over these slots. Let
$p_{u}\in\mathbb{C}^{n}$ be some signature from a given set $\mathcal{P}
\subset\mathbb{C}^{n}$ and $x_{u}\in\mathcal{X}^{n}$ be an unknown (uncoded)
data sequence from the modulation alphabet $\mathcal{X}^{n}$ both for the
$u$-th user with $u\in\{1,...,U\}$ and $U$ is the (fixed) maximum set of users
in the systems. Note that in our system $n$ is a very large number, e.g. 24k.
Due to the random zero-mean nature of $x_{u}$ we have $\frac{1}{n}E\lVert
p_{u}+x_{u}\rVert^{2}=1$, i.e. the total (normalized) transmit power is unity.
Provided user $u$ is active, we set:
\[
\alpha:=\frac{1}{n}\lVert p_{u}\rVert^{2}\quad\text{and}\quad\alpha^{\prime
}:=1-\alpha=\frac{1}{n}E\lVert x_{u}\rVert^{2}%
\]
Hence, the control signalling fraction of the power is $\alpha$. If a user is
not active then we set both $p_{u}=x_{u}=0$, i.e. either a user is active and
seeks to transmit data or it is inactive. Let $h_{u}\in\mathbb{C}^{s}$ denotes
the sampled channel impulse response (CIR) where $s\ll n$ is the length of the
cyclic prefix. The most important assumptions in this paper are:

\begin{itemize}
\item[(i)] Bounded support of $h_{u}$, i.e. $\text{supp}(h_{u})\subseteq
\lbrack0,\dots,s-1]$ due to the cyclic prefix

\item[(ii)] Sparsity of $h_{u}$ within supp$(h_{u})$, i.e. $\lVert h_{u}%
\rVert_{l_{0}}\leq k_{1}$

\item[(iii)] Sparse user activity, i.e. $k_{2}$ users out of $U$ in total are
actually active.
\end{itemize}

Define $k:=k_{1}k_{2}$.

Let $[h,0]\in\mathbb{C}^{n}$ denote the zero-padded CIR. The received signal
is then:
\begin{align*}
y  &  =\sum_{u=0}^{U-1}\text{circ}([h_{u},0])(p_{u}+x_{u})+e\\
y_{\mathcal{B}}  &  =\Phi_{\mathcal{B}}y
\end{align*}
Here, $\text{circ}([h_{u},0])\in\mathbb{C}^{n}$ is the circulant matrix with
$[h_{u},0]$ in its first column. $\Phi_{\mathcal{B}}$ denotes some measurement
matrix (to be specified later on) typically referring to a frequency window
$\mathcal{B}$ of size $m:=|\mathcal{B}|$. All performance indicators depend
strongly on the number of subcarriers in $\mathcal{B}$ (control) and
$\mathcal{B}^{C}$ (data). The goal is clearly a small observation window
$\mathcal{B}$.

The AWGN is denoted as $e\in\mathbb{C}^{n}$ with $E(ee^{\ast})=\sigma^{2}%
I_{n}$. For circular convolutions we have circ$([h,0])p=\sqrt{n}\cdot W^{\ast
}(\hat{h}\odot\hat{p})$ so that:
\begin{align*}
y  &  =\sum_{u=1}^{U}W^{\ast}\left[  (\sqrt{n}\hat{h}_{u}\odot(\hat{p}%
_{u}+\hat{x}_{u})\right]  )+\hat{e}\\
y_{\mathcal{B}}  &  =\Phi_{\mathcal{B}}y
\end{align*}
where $e$ and $\hat{e}$ are statistically equivalent.

\subsection{Control signaling model}

For the CCRA scheme let us assume that users' preambles 'live' entirely in
$\mathcal{B}$ while all data resides in $\mathcal{B}^{C}$, so that
supp$(p_{u})\subseteq\mathcal{B}\;\forall u$. We call this a common overloaded
control channel \cite{Wunder2015_GC}. Let $P_{\mathcal{B}}:\mathbb{C}%
^{n}\rightarrow\mathbb{C}^{m}$ be the corresponding projection matrix, i.e.
the submatrix of $I_{n}$ with rows in $\mathcal{B}$. For identifying which
preamble is in the system we can consider $\hat{y}$ and use the frequencies in
$\mathcal{B}$, i.e. $\Phi_{\mathcal{B}}=P_{\mathcal{B}}W$, so that:
\[
y_{\mathcal{B}}:=P_{\mathcal{B}}\sum_{u=1}^{U}\left[  \sqrt{n}\hat{h}_{u}%
\odot(\hat{p}_{u}+\hat{x}_{u})\right]  +P_{\mathcal{B}}\hat{e}%
\]
For algorithmic solution, we can stack the users as:%
\begin{align*}
y  &  =\sum_{u=1}^{U}\text{circ}(h_{u})(p_{u}+x_{u})+e\\
&  =D(p)h+C(h)x+e
\end{align*}
where $D(p):=[$circ$(p_{1}),\dots,$circ$(p_{U})]\in\mathbb{C}^{n\times Un}$
and $C(h):=[$circ$([h_{1},0]),\dots,$circ$([h_{U},0])]\in\mathbb{C}^{n\times
Un}$ are the corresponding compound matrices, respectively $p=[p_{1}%
^{T}\ p_{2}^{T}\ ...p_{U}^{T}]^{T}$ und $h=[h_{1}^{T}\ h_{2}^{T}\ ...h_{U}%
^{T}]^{T}$ are the corresponding compound vectors. If we assume each
user-channel vector $h_{u}$ to be $k_{1}$-sparse and $k_{2}$ are active then
$h$ is $k$-sparse.

For joint user activity detection and channel estimation exploiting the
sparsity we can use the standard basis pursuit denoising (BPDN) approach:%
\begin{equation}
\hbar=\arg\min_{h}\lVert h\rVert_{\ell_{1}}\;\text{s.t. }\lVert\Phi
_{\mathcal{B}}\,D(p)h-y\rVert_{\ell_{2}}\leq\epsilon\label{eqn:bpdn}%
\end{equation}
Moreover, several greedy methods such as CoSAMP exists for sparse
reconstruction. After running the algorithm in eqn. (\ref{eqn:bpdn}) the
decision variables $\lVert\hbar_{u}\rVert_{\ell_{2}}^{2}\;\forall u,$ are
formed, indicating that if $\lVert\hbar_{u}\rVert_{\ell_{2}}^{2}>\xi$ where
$\xi>0$ is some predefined threshold the user is considered active and its
corresponding data is detected.

\subsection{Data signaling model}

\begin{figure}[t]
\centering\includegraphics[width=1\linewidth]{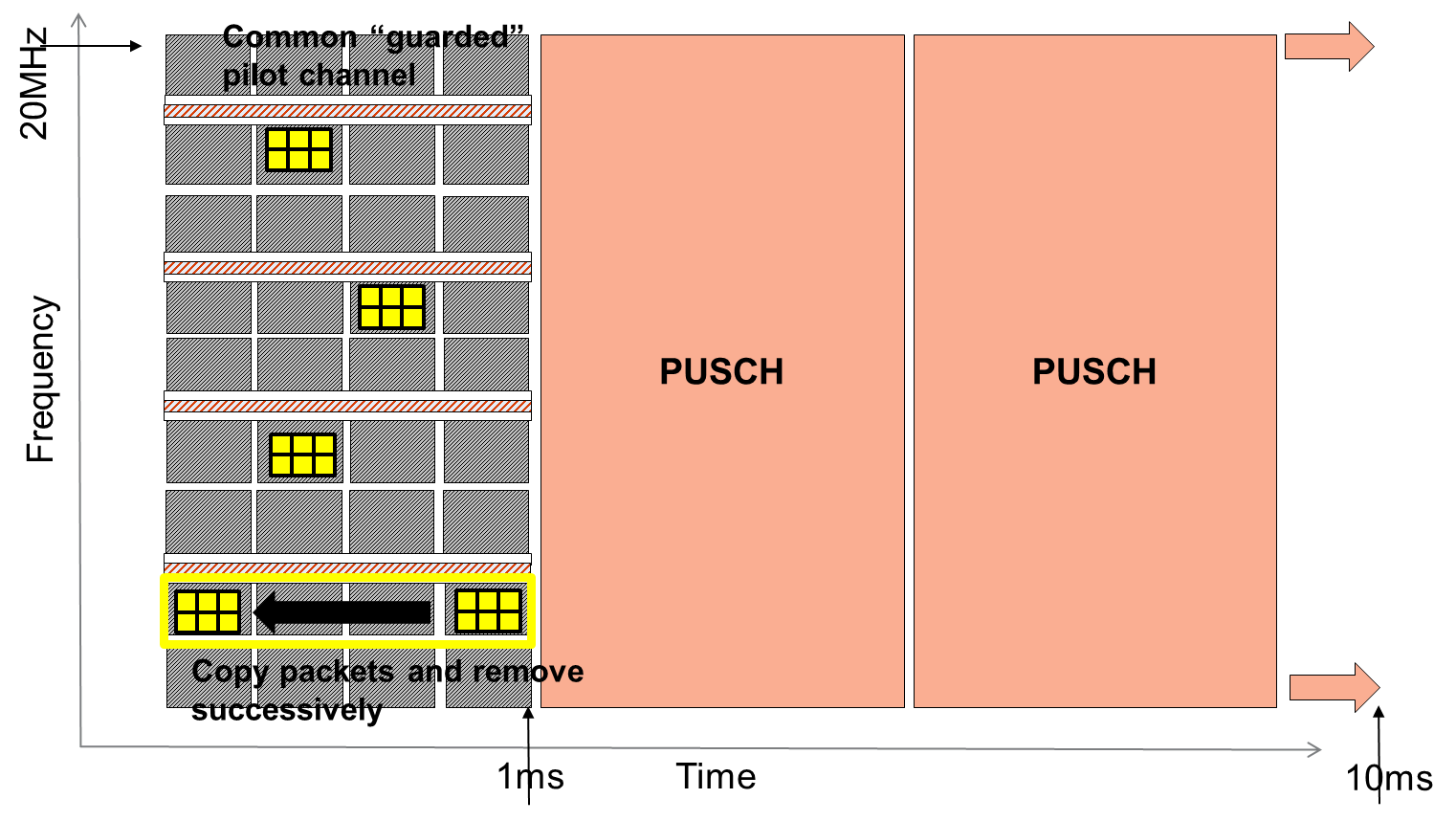}\caption{Random
access concepts: Approach.}%
\label{fig:approach}%
\end{figure}

Since data resides only in $\mathcal{B}^{C}$ the entire bandwidth
$\mathcal{B}^{C}$ can be divided into $B$ frequency patterns. Each pattern is
uniquely addressed by the preamble and indicates where the data and
corresponding copies are placed. the scheme works as follows: if a user wants
to transmit a small data portion, the pilot/data ratio $\alpha$ is fixed and a
preamble is randomly selected from the entire set. The signature determines
where (and how many of) the several copies in the $B$ available frequency
slots are placed which are processed in a specific way (see below). Such
copies can greatly increase the utilization and capacity of the traditional
ALOHA schemes. The principle is depicted in Fig. \ref{fig:approach}.

The random access algorithm can be seen as in instance of
\emph{coded slotted ALOHA} framework \cite{PSLP2015}, tuned to incorporate the
particularities of the physical layer addressed in the paper, as described in
the previous section. Specifically, the random access algorithms assumes that:

\begin{itemize}
\item the users are active in multiple \CS{frequency} slots,
denoted simply as slots in further text,

\item the activity pattern, i.e., the choice of the slots is random, according
to a predefined distribution,

\item every time a user is active, it sends a replica of packet, which
contains data,

\end{itemize}

Obviously, due to the random nature of the choice of slots, the access point
(i.e. the base station) observes idle slots (with no active user), singleton
slots (with a single active user) and collision slots (with multiple active
users). Using a compressive sensing receiver, the base station, decodes
individual users from non-idle slots,
removes (cancels) the replicas from the slots in which they occur \CS{(the knowledge of which is learned through signatures)}, and tries to decode new users from the slots from
which replicas (i.e. interfering users) have been cancelled. In this way, due
to the cancelling of replicas, the slots containing collisions that previously
may have not been decodable, can become decodable. This process is executed in
iterations, until there are no slots from which new users can be decoded. The
above described operation can be represented via graph, see
Fig.~\ref{fig:CSA_example}.

\begin{figure2}
[t]
\centering\includegraphics[width=1.7\columnwidth]{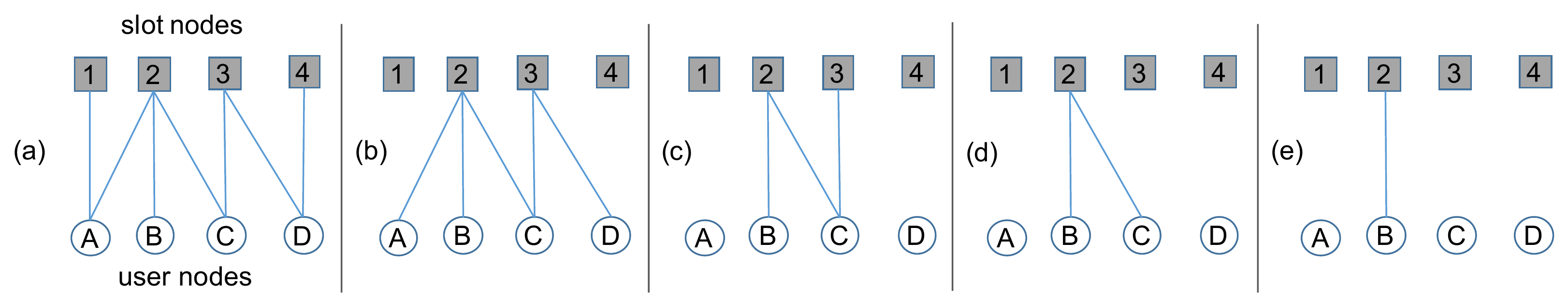}\caption{Example
of iterative IC decoding; it is assumed that only singleton slots are
decodable, with probability 1. (a) The base stations decodes singleton slots 1
and 4, (b) obtaining packets of users A and D. (c) In the next step, the base
station cancels replicas of decoded packets from slots 2 and 3, respectively,
reducing slot 3 to a singleton slot. (d) The base station decodes packet of
user C, and (e) cancels its replica from slot 2, which now becomes singleton.}\label{fig:CSA_example}%

\end{figure2}

The iterative interference cancellation (IC) resembles iterative
belief-propagation erasure decoding, allowing the use of the related
theoretical concepts to analyze and design random access algorithms. However,
the important differences have to be taken account, stemming from the nature
of the physical layer operation:

\begin{itemize}
\item[(i)] The received singleton slots are not always decodable, i.e. they
are decodable with a certain probability, which depends on the received SNR,
channel estimation etc.

\item[(ii)] \CS{The received collision slots may be decodable, depending on the
multi-user detection capabilities. Further, the unbalance of received signal
powers due to varying channels that users experience, may cause capture
effect, where a subset of the collided users may be decoded as a result of a
favorable SINR}.

\item[(iii)] Cancellation of replicas, in general, is not ideal, due to
imperfect channel estimation and/or channel variations among the slots where
replicas occurred, operation of the physical layer etc., and leaves a residual
interference power. This implies that, as the IC progresses, the residual
interference accumulates in the affected slots, which may prevent further
decoding of the remaining user packets.
\end{itemize}

Analytical modeling of the above is the main prerequisite to assess the
performance of the random access algorithm, which in turn, allows for the
design of the probability distribution that governs the choice slots, and
which is typically optimized to maximize the throughput, i.e., the number of
resolved packets per slot \cite{PSLP2015}.

In the context of application of coded slotted ALOHA to compressive-sensing
based physical layer, some preliminary work can be found in \cite{Popovski14}.
Here we extend the approach, by taking into account a more detailed operation
of the physical layer, which incorporates channel estimation and imperfect
interference cancellation, as detailed in Section~\ref{sec:CSA}.

\section{Performance analysis}

The performance analysis is split into activity detection/channel estimation
and the data part, where coded random access is included.

\subsection{User detection/channel estimation}

In the data model we assume that fast fading effects are averaged out due to
coding over subcarriers. Hence, user rates are ergodic and are calculated as
expectations over the fading distributions. Achievable rates crucially depend
on the receive powers (user position, slow fading effects), channel estimation
errors and corresponding interference from colliding users then
\cite{Wunder2015_GC}. The relevant expressions under erroneous channel
estimation will be provided below.

Suppose user $u$ as well as colliding users $u\left(  j\right)  \in
\mathcal{C}_{u},j=1,...|\mathcal{C}|$, which are detected before in some
singleton slot have been assigned subcarriers $i\in\mathcal{B}_{u}$. Due to
the circular model each subcarrier has powers $E(|\hat{x}_{u,i}|^{2}%
)=1-\alpha$, $|\hat{p}_{u,i}|^{2}=\alpha$ and $E(|\hat{e}_{u,i}|^{2}%
)=\sigma^{2}$. Denote the channel estimation error as $\hat{d}_{u,i}%
:=\hat{\hbar}_{u,i}-\hat{h}_{u,i}$. Hence, the received signal is given by:
\[
\hat{y}_{u,i}=(\sqrt{n}\hat{\hbar}_{u,i}+\hat{d}_{u,i})\hat{x}_{u,i}+\hat
{e}_{u,i}%
\]
for singleton slots and
\[
\hat{y}_{u,i}=(\sqrt{n}\hat{\hbar}_{u,i}+\hat{d}_{u,i})\hat{x}_{u,i}%
+\sum_{j\in\mathcal{C}}\hat{d}_{u\left(  j\right)  ,i}\hat{x}_{u\left(
j\right)  ,i}+\hat{e}_{u,i}%
\]
for collision slots. Suppose further we have calculated the propability of not
detecting an active user $P_{md}(\xi)$ ("missed detection"), and falsely
detecting an inactive user $P_{fa}(\xi)$ ("false alarm"). Define $\bar{P}%
_{md}(\xi):=1-P_{md}(\xi)$ \cite{Wunder2015_GC}. Let the channel impulse
response be $k$-sparse and use BPDN as the channel estimate. Further, let
$\Phi_{\mathcal{B}},m=|\mathcal{B}|$, be a fixed measurement matrix with RIP
constant $\delta_{2k}<\sqrt{2}-1$ and corresponding $c_{1}(\delta_{2k})$. The
achievable rate $R(\alpha)$ per subcarrier for a particular user is lower bounded

\begin{itemize}
\item for singleton slots by:%
\begin{align*}
R\left(  \alpha\right)   &  \geq E_{h|\{\left\Vert h\right\Vert >\xi\}}\left[
\log\left(  1+\frac{\left(  1-\alpha\right)  |h|^{2}}{\sigma^{2}}\right)
\right]  \bar{P}_{md}(\xi)\\
&  -\log\left(  1+\frac{\left(  1-\alpha\right)  c_{1}(\delta_{2k})^{2}%
m}{\sigma^{2}\alpha nk_{2}}\right)
\end{align*}

\item and for collisions slots by:%
\begin{align*}
R\left(  \alpha\right)   &  \geq E_{h|\{\left\Vert h\right\Vert >\xi\}}\left[
\log\left(  1+\frac{\left(  1-\alpha\right)  |h|^{2}}{\sigma^{2}}\right)
\right]  \bar{P}_{md}(\xi)\\
&  -\log\left(  1+\frac{\left(  |\mathcal{C}|+1\right)  \left(  1-\alpha
\right)  c_{1}(\delta_{2k})^{2}m}{\sigma^{2}\alpha nk_{2}}\right)
\end{align*}

\end{itemize}

To prove we can extend the analysis in \cite{Wunder2015_GC} in a
straightforward manner. Note that the performance strongly depends on the
scaling of $\frac{c_{1}(\delta_{2k})^{2}m}{nk_{2}}$. From the CS literature
upper and lower bounds are available (e.g. for CoSAMP see \cite{Foucart2012}),
i.e. $c_{1}(\delta_{2k})=4\sqrt{1+\delta_{2k}}/1-(1+\sqrt{2})\delta_{2k}$ as
well as bounds on the RIP constants $\delta_{2k}$ \cite{Rudelson:2007}, but
these bounds are rather loose so numerial simulations are still necessary.

\subsection{Coded Slotted Aloha}

\label{sec:CSA}

\begin{figure}[t]
\centering\includegraphics[width=0.9\columnwidth]{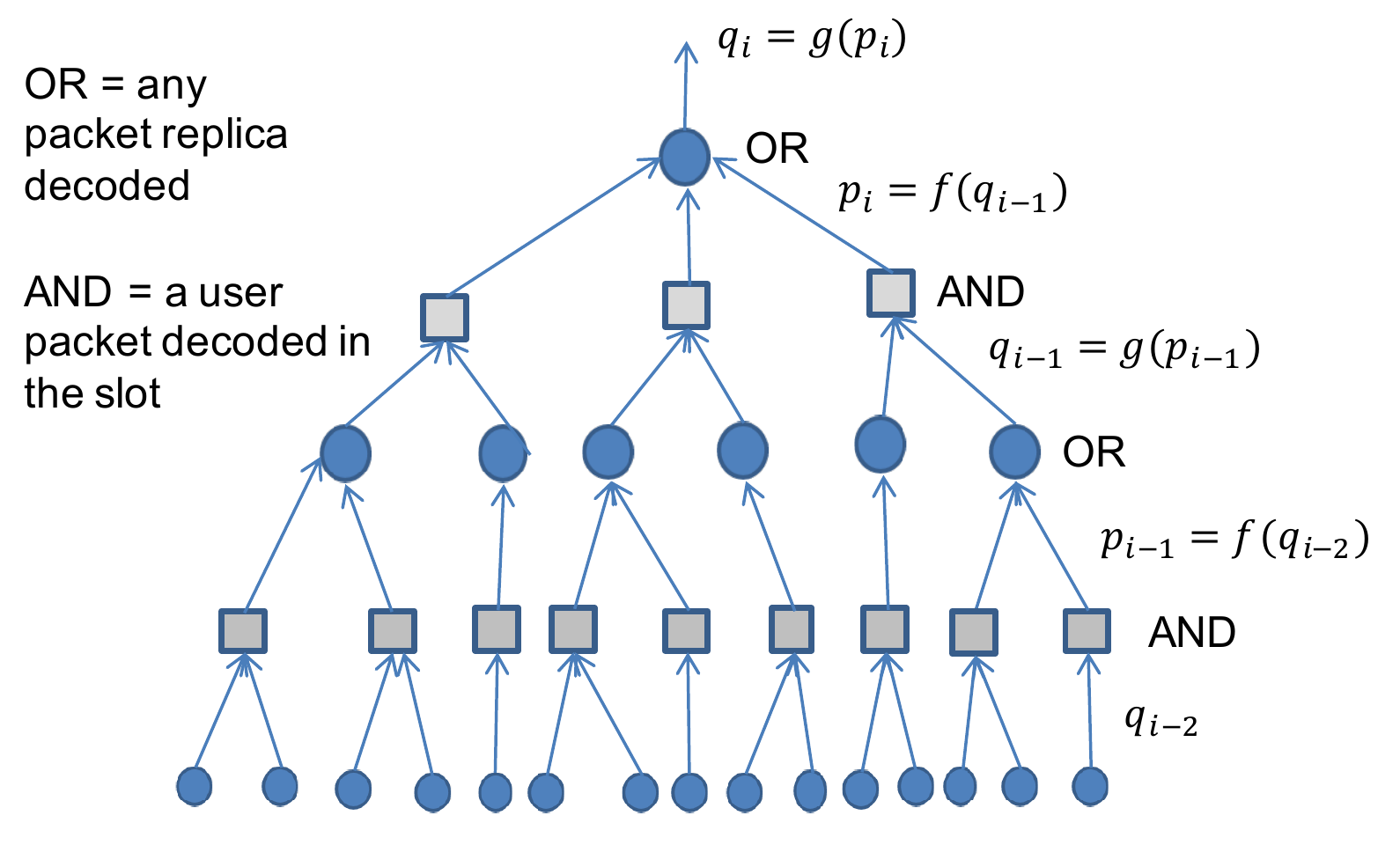}\caption{And-or tree
evaluation.}%
\label{fig:tree}%
\end{figure}

The analysis of coded slotted ALOHA is typically based on the and-or tree
evaluation \cite{LMS1998}. It is assumed that the graph representation can be
unfolded in a tree, see Fig.~\ref{fig:tree}, on which two operations are
performed in succession:

\begin{itemize}
\item[(i)] decoding of user packets in slots, corresponding to (a generalized)
``and'' operation, cf. \cite{SMP2014,Popovski14},

\item[(ii)] removal of replicas, corresponding to ``or'' operation.
\end{itemize}

Both operations are probabilistically characterized, in terms of probability
of \emph{not} decoding a user in a slot, denoted as $p_{i}$, and \emph{not}
removing a replica $q_{i}$. The tree structure allows for their successive
updates, as reflected in the subscripts of $p_{i}$ and $q_{i}$.
We also note that the analysis is asymptotic in nature, as in the
non-asymptotic case, the graph representation contains loops, and the
corresponding tree representation is only an approximation.

Before giving providing the expressions for $p_{i}$ and $q_{i}$, we introduce
the following terminology. Denote the number of edges incident to slot/user
node as slot/user node degree. Further, by edge-oriented slot degree
distribution $\omega_{j}$, $j\geq1$ and $\sum_{j}\omega_{j}=1$, denote the
probability distribution that a randomly chosen edge in the graph is connected
to a slot node of degree $j$ \cite{LMS1998}. Similarly, by edge-oriented user
degree distribution $\lambda_{k}$, $k\geq1$ and $\sum_{k}\lambda_{k}=1$,
denote the probability distribution that a randomly chosen edge in the graph
is connected to a user node of degree $j$ \cite{LMS1998}. Note that
$\lambda_{k}$ are subject to design of the random access algorithm, and that
they implicitly determine $\omega_{j}$. It could be shown that the probability
update in slot node is:
\begin{equation}
p_{i}=\sum_{j}\omega_{j}\sum_{t=0}^{j-1}\pi_{t,j}{\binom{j-1}{t}}q_{i-1}%
^{t}(1-q_{i-1})^{j-t-1},\;i\geq1, \label{eqn:p_i}%
\end{equation}
where $j$ is the slot degree, $t$ is the number of interfering users that
decreases through iterations via use of IC, $\pi_{t,j}$ is the probability of
decoding a user packet in the slot of degree $j$ when $t$ interfering packets
have been cancelled, and where the combinatorial term ${\binom{j-1}{t}}$ stems
from the assumption that all colliding user packets in the slot are
statistically a-priori the same, in terms of probability of being
decoded\footnote{Eqn. \eqref{eqn:p_i} can be derived using the approach
similar to \cite{SMP2014,Popovski14}.}. Here is important to note that the
direct influence of the physical layer, i.e., receiver operation, as described
in \ref{sec:CCRA}, is embedded in $\pi_{t,j}$. The probability update in user
node is:
\[
q_{i}=\sum_{k}\,\lambda_{k}p_{i}^{k-1},\;i\geq1,
\]
with the initial value $q_{0}=1$. Finally, the output of the evaluation is the
probability that a user packet is decoded:
\[
P_{D} = 1 - \lim_{i \rightarrow\infty}q_{i}.
\]

\section{Simulations}

An LTE-A 4G frame consists of a number of subframes with 20MHz bandwidth; the
first subframe contains the RACH with one "big" OFDM symbol of 839 dimensions
located around the frequency center of the subframe. The FFT size is 24578=24k
corresponding to the 20MHz bandwidth whereby the remainder bandwidth outside
PRACH is used for scheduled transmission in LTE-A, so-called PUSCH. The prefix
of the OFDM symbol accommodates delays up to 100$\mu s$ (or 30km cell radius)
which equals 3000 dimensions. In the standard the RACH is responsible for user
aquisition by correlating the received signal with preambles from a given set.
Here, to mimic a 5G situation, we equip the transmitter with the capability of
sending information in "one shot", i.e., in addition to user aquisition,
channel estimation is performed and the data is detected. For this a fraction
of the PUSCH is reserved for data packets of users which are detected in the
PRACH. Please note the rather challenging scenario of only 839 subcarrier in
the measurement window versus almost 24k data payload subcarriers.

In our setting, a limited number of users is detected out of a maximum set
(here 10 out of 100). We assume that the delay spread is below 300 dimensions
of which only a set of 6 pathes are actually relevant. The pilot signalling is
similar to \cite{Wunder2014_ICC} but modifed to fit the data/pilot separation.

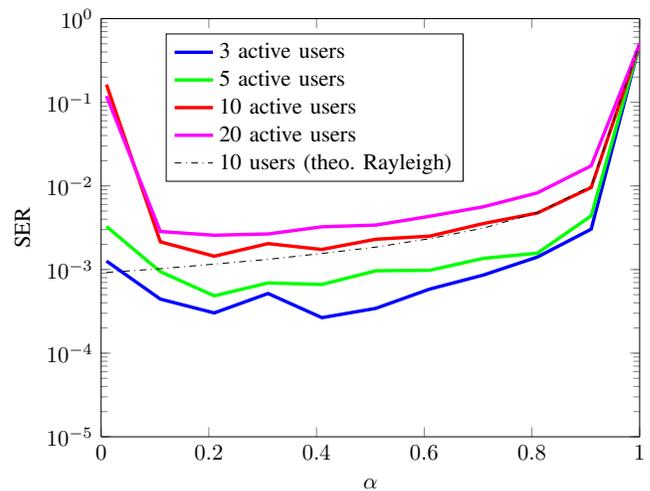
\begin{figure}[t]
\centering
\definecolor{mycolor1}{rgb}{1.00000,0.00000,1.00000}\begin{tikzpicture}[scale=.85]
\begin{axis}[%
width=0.951\linewidth,
height=0.739\linewidth,
at={(0\linewidth,0\linewidth)},
scale only axis,
separate axis lines,
every outer x axis line/.append style={black},
every x tick label/.append style={font=\color{black}},
xmin=0,
xmax=1,
xlabel={$\alpha$},
every outer y axis line/.append style={black},
every y tick label/.append style={font=\color{black}},
ymode=log,
ymin=1e-05,
ymax=1,
yminorticks=true,
ylabel={SER},
axis background/.style={fill=white},
legend style={at={(0.12,0.61)},anchor=south west,legend cell align=left,align=left,fill=none,draw=black}
]
\addplot [color=blue,solid,line width=1.5pt]
table[row sep=crcr]{%
0.01	0.00126256630627449\\
0.11	0.00044303383500448\\
0.21	0.000302637861455043\\
0.31	0.000516845554618056\\
0.41	0.000265633891254439\\
0.51	0.000343181981699141\\
0.61	0.000583168180837078\\
0.71	0.000859716853403449\\
0.81	0.00140950297274493\\
0.91	0.00303708899920858\\
1	0.498041176932284\\
};
\addlegendentry{3 active users};
\addplot [color=green,solid,line width=1.5pt]
table[row sep=crcr]{%
0.01	0.00327307147669224\\
0.11	0.000942970880835116\\
0.21	0.000484672703726233\\
0.31	0.000691702803995055\\
0.41	0.000662993250004905\\
0.51	0.00096603172890136\\
0.61	0.000979296254145164\\
0.71	0.0013612277534682\\
0.81	0.0015669431744599\\
0.91	0.00438431803465259\\
1	0.500378117457764\\
};
\addlegendentry{5 active users};
\addplot [color=red,solid,line width=1.5pt]
table[row sep=crcr]{%
0.01	0.162856271652605\\
0.11	0.00214429368147261\\
0.21	0.0014421791844628\\
0.31	0.00204036534130426\\
0.41	0.00173815971063975\\
0.51	0.00230231791255966\\
0.61	0.00250683447904471\\
0.71	0.00354922852526997\\
0.81	0.0047368799150582\\
0.91	0.00955871212121212\\
1	0.500524364626189\\
};
\addlegendentry{10 active users};
\addplot [color=mycolor1,solid,line width=1.5pt]
table[row sep=crcr]{%
0.01	0.118943622374555\\
0.11	0.00284773190034843\\
0.21	0.0025717851094892\\
0.31	0.00266344961125937\\
0.41	0.00324302167751394\\
0.51	0.00340034890779122\\
0.61	0.0043270061960958\\
0.71	0.00563269283362522\\
0.81	0.0082474562930406\\
0.91	0.0173883783471048\\
1	0.499858851822\\
};
\addlegendentry{20 active users};
\addplot [color=black,dashdotted]
table[row sep=crcr]{%
0.01	0.000923194948055817\\
0.11	0.00102272521043062\\
0.21	0.00116047959743498\\
0.31	0.00132304074491651\\
0.41	0.00155682077477703\\
0.51	0.00185322203098892\\
0.61	0.00234067111650649\\
0.71	0.00313174194111054\\
0.81	0.00478103578180689\\
0.91	0.00995635573568282\\
1	0.499999999999995\\
};
\addlegendentry{10 users (theo. Rayleigh)};
\end{axis}
\end{tikzpicture}
\caption{Averaged BPSK SER in 5G \textquotedblright one-shot\textquotedblright%
\ random access in a $20$MHz LTE-A standard setting at (overall) SNR=$20$dB.
$m=839$ out of $n=24576$ dimensions are used for CS and sparsity of the
channel is $k=6$. The total number of users is 100, out of which 3-20 are
active. The control overhead is below $13$\%.}%
\label{fig:crach:ser}%
\end{figure}

\section{Conclusions}

In this paper, we provided ideas how to enable random access for many devices
in a massive machine-type scenario. In the conceptional approach as well as
the actual algorithms sparsity of user activity and channel impulse responses
plays an a pivotal role. We showed that using such framework efficient "one
shot" random access is possible where users can send a message without a
priori synchronizing with the network. Key is a common overloaded control
channel which is used to jointly detect sparse user activity and sparse
channel profiles. Such common control channel stands in clear contrast to
dedicated control signalling per ressource block, and is thus more efficent
particularly for small ressoure blocks. Since each user also has channel state
information for all subcarriers, there are additional degrees of freedom to
place the ressource blocks. We analyzed the system theoretically and provided
a link between achievable rates and standard compressing sensing estimates in
terms of explicit expressions and scaling laws. Finally, we supported our
findings with simulations in an LTE-A-like setting allowing "one shot" sparse
random access of 100 users in 1ms with good performance.

\bibliographystyle{IEEEtran}


\end{document}